\documentclass[12pt]{article}
\usepackage[cp866]{inputenc}
\usepackage[english]{babel}
\usepackage{epsfig}
\usepackage{amsbsy}
\usepackage{amsfonts}

\begin{document}
\title{
Hyperbolic attractor in a system of coupled non-autonomous van der Pol 
oscillators: Numerical test for expanding and contracting cones
}

\author{S.P. Kuznetsov and I.R. Sataev}
\date{}

\maketitle

\begin{center}
Institute of Radio-Engineering and Electronics 
of RAS, Saratov Branch,\\ Zelenaya 38, 410019, Saratov, Russian Federation
\end{center}

\begin{abstract}
We present numerical verification of hyperbolic nature for chaotic attractor 
in a system of two coupled non-autonomous van der Pol oscillators 
(Kuznetsov, Phys. Rev. Lett., \textbf{95, }144101, 2005). At certain 
parameter values, in the four-dimensional phase space of the Poincar\'{e} 
map a toroidal domain (a direct product of a circle and a three-dimensional 
ball) is determined, which is mapped into itself and contains the attractor 
we analyze. In accordance with the computations, in this absorbing domain 
the conditions of hyperbolicity are valid, which are formulated in terms of 
contracting and expanding cones in the tangent spaces (the vector spaces of 
the small state perturbations). 
\end{abstract}

Mathematical theory of chaotic dynamics based on a rigorous axiomatic 
foundation exploits a concept of hyperbolicity [1-8]. 

An orbit in phase space of a dynamical system is called hyperbolic if there 
are trajectories approaching exponentially the original orbit, and those 
departing from it in a similar manner. Moreover, an arbitrary small 
perturbation of a state on the original orbit must admit representation via 
a linear combination of the growing and the decaying perturbations. 

In dissipative systems contracting the space volume the attractors may 
occur, which consist exclusively of the hyperbolic orbits. These are 
attractors with strong chaotic properties, like existence of the 
well-defined invariant SRB-measure, a possibility of description in terms of 
Markov partitions and symbolic dynamics, positive metric and topological 
entropy etc. Such hyperbolic (or, more definitely, \textit{uniformly hyperbolic}) attractors are robust 
or structurally stable, that means insensitivity of the type of dynamics and 
of the phase space structure in respect to slight variations of functions 
and parameters in the evolutionary equations. 

Although the basic statements of the hyperbolic theory were formulated 40 
years ago, no convincing examples of physical systems were introduced with 
uniform hyperbolic attractors. In textbooks and reviews on nonlinear 
dynamics, such attractors are represented by artificial mathematical 
constructions, like Plykin attractor and Smale -- Williams solenoid [1-8]. 
For realistic systems, in which the chaotic dynamics is mathematically 
proved, like the Lorenz model [9,10], the strange attractors do not relate 
to the class of uniformly hyperbolic (not all axiomatic statements of the 
classic hyperbolic theory are valid for them). Some aspects of possible 
existence of hyperbolic attractors in differential equations were discussed 
e.g. in Refs. [11-14]. 

In a recent paper of one of the authors [15], an idea was advanced of 
implementation of a hyperbolic attractor in a system of two coupled 
non-autonomous van der Pol oscillators. In a Poincar\'{e} map that 
determines evolution on a period of the external driving, a chaotic 
attractor has been found, which demonstrates some characteristic signs of 
hyperbolic attractors. By a nature of transformation of the phase space 
volume in a course of the evolution over a period, it is similar to the 
Smale -- Williams solenoid. It looks robust: the Cantor-like transverse 
structure and the positive Lyapunov exponent are insensitive to variation of 
parameters in the equations. An analogous system has been built as an 
electronic device and studied in experiment [16].

Obviously, it would be desirable to have a mathematical confirmation of the 
hyperbolic nature of the attractor. As Sinai has suggested in due time [1], 
one possible way for substantiation of the hyperbolicity for attractor of a 
Poincar\'{e} map consists in numerical verification of certain sufficient 
conditions formulated in terms of inclusion for expanding and contracting 
cones in tangent vector space (the space of small perturbation vectors). In 
this paper, we discuss a method and present results of computer verification 
of these conditions in application to the chaotic attractor in a system of 
two coupled non-autonomous van der Pol oscillators. 

The system proposed in Ref. [15] is represented by a set of differential 
equations 
\begin{equation}
\label{eq1}
\begin{array}{ll}
 \dot {x} = \omega _0 u, &
 \dot {u} = - \omega _0 x + (A\cos {2\pi t/T} - x^2)u 
+ (\varepsilon / {\omega _0 })y\cos \omega _0 t, \\
 \dot {y} = 2\omega _0 v, &
 \dot {v} = - 2\omega _0 y + ( - A\cos {2\pi t/T} - y^2)v 
+ (\varepsilon / {2\omega _0 })x^2. \\
 \end{array}
\end{equation}
It consists of two subsystems, the van der Pol oscillators with 
characteristic frequencies $\omega _{0}$ and 2$\omega _{0}$. Here $x$ and 
$u$ represent coordinate and velocity for the first oscillator, and $y$ and $v$ for 
the second one. In each oscillator the parameter responsible for the birth 
of the limit cycle, is forced to swing slowly with period $T$ and amplitude $A$. As 
the parameter modulation is of opposite phase, the subsystems generate turn 
by turn, each on its own half-period $T$. The coupling is characterized by 
parameter $\varepsilon$. The first oscillator affects the second one via a 
quadratic term in the equation. The backward coupling is introduced by a 
product of the variable $y$ and an auxiliary signal of frequency 
$\omega_{0}$. It is assumed that the interval 
$T$ contains an integer number of 
periods of the auxiliary signal 
$N_0 = {\omega _0 T}/{2\pi}$, so the external driving is periodic. 
For a detailed study, we select 
\begin{equation}
\label{eq2}
\omega _0 = 2\pi ,\,\,T = 6,\,\,A = 5,\,\,\varepsilon = 0.5.
\end{equation}

Qualitatively, the system (\ref{eq1}) operates as follows. 
Let the first oscillator 
on a stage of generation have some phase 
$\psi $: $x \propto \sin (\omega _0 t + \psi )$. 
The squared value $x^{2}$ contains the second harmonic: 
$\cos (2\omega _0 t + 2\psi )$, and its phase is 2$\psi $. As the half-period 
comes to the end, the term $x^{2}$ effects as priming for the excitation of 
the second oscillator, and the oscillations of $y$ get the phase 2$\psi$. Half 
a period later, the mixture of these oscillations with the auxiliary signal 
stimulates excitation of the first oscillator, which accepts this phase 
2$\psi$. Obviously, on subsequent periods the phase of the first oscillator 
will follow approximately the relation
\begin{equation}
\label{eq3}
\psi _{n + 1} = 2\psi _n + \mbox{const}\,\,\,(\bmod 2\pi ).
\end{equation}
(Here the constant accounts a phase shift in a course of transfer of the 
excitation from one oscillator to another and back.) 
The relation (\ref{eq3}) called 
the Bernoulli map is well known as one of the simplest 
model examples in the 
chaos theory.\footnote{The constant in Eq. (\ref{eq3}) 
may be removed by a shift of 
origin for the phase variable. We stress that the phase $\psi$ cannot be 
defined globally on the whole time interval $T$: it has sense only in the 
context of the discrete time description. Indeed, on the stage when the 
first oscillator does not generate, its amplitude is small, and phase is not 
well defined.} 

For accurate description of the discrete time dynamics, we turn to the 
Poincar\'{e} map [2-8, 17,18]. Let us have a vector 
${\rm {\bf x}}_n = \{x(t_n ),u(t_n ),y(t_n ),v(t_n )\}$ 
as a state of the system at 
$t_{n}$=\textit{nT}. From solution of the differential equations 
(\ref{eq1}) with the initial 
condition ${\rm {\bf x}}_n$, we get a new vector ${\rm {\bf x}}_{n + 1} $ 
at $t_{n + 1}$=($n$+1)$T$. As it is determined uniquely by 
${\rm {\bf x}}_n $, we introduce a function that maps 
the four-dimensional space 
$\{x,u,y,v\}$ into itself: 
${\rm {\bf x}}_{n + 1} = {\rm {\bf T}}({\rm {\bf x}}_n )$. 

This Poincar\'{e} map appears due to evolution determined by 
differential equations with smooth and bounded right-hand parts in a finite 
domain of variables $\{x,u,y,v\}$. In accordance with theorems of existence, 
uniqueness, continuity, and differentiability of solutions of differential 
equations, the map \textbf{T} is a diffeomorphism, a one-to-one 
differentiable map of class $C^{\infty }$ [17].

Further, we will deal always with description of the dynamics in terms of 
the Poincar\'{e} map. In particular, under the phase space we mean the 
four-dimensional space $\{x,u,y,v\}$, with $x$, $u$, $y$, $v$ 
relating to an instant 
$t_{n}$. An orbit means a discrete sequence of points in this space; 
attractor is an invariant attractive set composed of such orbits etc. 

In a course of iterations of the map 
${\rm {\bf x}}_{n + 1} = {\rm {\bf T}}({\rm {\bf x}}_n )$, 
we have expansion of a small phase-space volume 
in a direction associated with the phase 
in the approximate equation (\ref{eq3}) and 
contraction in the rest three directions. Interpreting the mapping 
geometrically, let us imagine a solid toroid embedded in the 4-dimensional 
space (a direct product of a circle and a three-dimensional ball) and 
associate one iteration of the map with longitudinal stretch of the toroid, 
with contraction in the transversal directions, and insertion of the doubly 
folded ``tube'' into the original toroid. It is analogous to the 
construction of Smale and Williams with the only difference that we deal 
with four-dimensional rather than the three-dimensional phase space.

The mentioned toroid will be referred to as an absorbing domain $U$. It means 
that under application of the map \textbf{T} the images of all points from 
$U$ belong to its interior: 
${\rm {\bf T}}(U) \subset \mbox{Int}\,U$. 
The attractor may be defined as intersection of the images of the original 
domain under multiple action of the map: 
$A = \bigcap\limits_{n = 1}^\infty {{\rm {\bf T}}^n(U)}$. 

To write down an analytic expression for the domain $U$ it is convenient to 
redefine the coordinate system. We introduce new variables 
$\{ x_{0},\, x_{1},\, x_{2},\, x_{3} \}$ as follows:
\begin{equation}
\label{eq4}
x_0 = x / r_0 ,\,\,\,\,x_1 = (u - c_{ux} x) / r_1 ,\,\,\,x_2 = y - c_{yx} x 
- c_{yu} u,\,\,\,x_3 = v - c_{vx} x - c_{vu} u - c_{vy} y.
\end{equation}

To determine the constants  
we accumulate a large number of 
points $\{x,u,y,v\}$ on the attractor in the Poincar\'{e} section by 
numerical solution of the equations (\ref{eq1}). Then, by the 
least square method we 
find out the coefficients to minimize the mean-square values 
$<(u - c_{ux} x)^2 >$, $<(y - c_{yx} x - c_{yu} u)^2 >$, 
$<(v - c_{vx} x - c_{vu} u - c_{vy} y)^2>$. 
Geometrically, it corresponds to directing the coordinate 
axes along the principal axes of ellipsoid that approximates the attractor. 
Additionally, we normalize $x_{0}$ and $x_{1}$ by appropriate factors to have 
<$x_0^2 $>=<$x_1^2 $>$ \approx $1/2. Finally, at the parameter set (\ref{eq2}) we 
get
\begin{equation}
\label{eq5}
\begin{array}{l}
 c_{ux} = 0.438,\,\,c_{yx} = - 0.042,\,\,\,c_{yu} = 0.226,\,\,c_{vx} = - 
0.218,\,\,\, \\ 
 c_{vu} = 0.029,\,\,c_{vy} = - 0.118,\,\,\,\,r_0 = 0.812,\,\,\,\,\,\,r_1 = 
0.721. \\ 
 \end{array}
\end{equation}

In the new coordinates, let us define the absorbing domain $U$ by the 
inequality:
\begin{equation}
\label{eq6}
\left[ {\left( {\sqrt {x_0^2 + x_1^2 } - r} \right) / d_r } \right]^2 + 
\left( {x_2 / d} \right)^2 + \left( {x_3 / d} \right)^2 \le 1.
\end{equation}
Empirically selected constants in this expression are 
$r = 0.94,\,\,d_r = 0.4,\,\,d = 0.15$. 
Figure 1 gives evidence that this is indeed an absorbing 
domain. For initial points distributed over a border of $U$ we perform 
numerical solution of the differential equations on an interval $T$ and plot 
the results in the coordinates 
\begin{equation}
\label{eq7}
R_1 = \left( {\sqrt {x_0^2 + x_1^2 } - r} \right) / d_r ,\,\,\,
R_2 = \sqrt {\left( {x_2 / d} \right)^2 + \left( {x_3 / d} \right)^2} .
\end{equation}
As the whole figure is placed inside the unit circle $R_1^2 + R_2^2 = 1$, 
the images of the initial points belong to the interior of $U$. 

In Fig.2 we show a three-dimensional projection to illustrate mutual 
location of the domains $U$ and \textbf{T}$(U)$. It is analogous to that 
considered on the first step of the construction of the Smale -- Williams 
attractor: take a torus (``a plastic doughnut''), stretch it twice, contract 
transversally, fold twice and squeeze into its original volume. The 
transformed ``doughnut'' \textbf{T}$(U)$ looks like a narrow band because of 
very strong compression of the phase volume in respective directions in a 
course of the evolution.

We will verify hyperbolicity conditions required by a theorem 
(see e.g. [1, 13]) adopted for the problem under consideration. 
Unlike the general 
formulation, it is sufficient for us to deal with a diffeomorphism of class 
C$^{\infty }$ in the Euclidian space 
${\rm \mathbb{R}}^{4}$ 
$\{ x_{0},\, x_{1},\, x_{2},\, x_{3} \}$. 
That is the Poincar\'{e} map \textbf{T}(\textbf{x}). 
Evolution of a perturbed state ${\rm {\bf x}} + \delta {\rm {\bf x}}$ 
corresponds to transformation of the perturbation vector 
$\delta {\rm {\bf x}}$ in linear approximation 
$\delta {\rm {\bf x}}' = {\rm {\bf DT}}_{\rm {\bf x}} \delta {\rm {\bf x}}$, 
where 
${\rm {\bf DT}}_{\rm {\bf x}} $ 
is the Jacobi matrix at \textbf{x}: 
${\rm {\bf DT}}_x = \{\partial {x}'_i / \partial x_j \},\,\,i,j = 0,\,1,\,2,\,3.$
The notion ${\rm {\bf DT}}_{\rm {\bf x}}^{ - 1}$ 
designates the derivative matrix for the 
inverse mapping ${\rm {\bf T}}^{ - 1}({\rm {\bf x}})$. 

\textbf{Theorem [1,13].} Suppose that a diffeomorphism \textbf{T} of class 
C$^{\infty }$ maps a bounded domain $U \subset {\rm \mathbb{R}}^{4}$ 
into itself: 
${\rm {\bf T}}(U) \subset \mbox{Int}\,U$, and $A \subset \mbox{Int}\,U$ is 
an invariant subset for the diffeomorphism. The set $A$ will be uniformly 
hyperbolic if there exists a constant $\gamma >1$ and the following 
conditions hold:

\begin{enumerate}
\item
For each ${\rm {\bf x}} \in A$ in the space $\mathbb{V} _{x}$ of 4D vectors 
$\delta {\rm {\bf x}}$ the expanding and contracting cones 
$S_{\rm {\bf x}}^\gamma $ and $C_{\rm {\bf x}}^\gamma $ 
may be defined, such that 
$\left\| {{\rm {\bf DT}}_{\rm {\bf x}} {\rm {\bf u}}} \right\| > \gamma 
\left\| {\rm {\bf u}} \right\|$ for all ${\rm {\bf u}} \in S_{\rm {\bf 
x}}^\gamma $, 
and 
$\left\| {{\rm {\bf DT}}_{\rm {\bf x}}^{ - 1} {\rm {\bf 
v}}} \right\| > \gamma \left\| {\rm {\bf v}} \right\|$ for all ${\rm {\bf 
v}} \in C_{\rm {\bf x}}^\gamma $; 
moreover, for all ${\rm {\bf x}} \in A$ 
they satisfy 
$S_{\rm {\bf x}}^\gamma \cap C_{\rm {\bf x}}^\gamma = \emptyset 
$ and $S_{\rm {\bf x}}^\gamma + C_{\rm {\bf x}}^\gamma = \mathbb{V} _{x}$.

\item
The cones $S_{\rm {\bf x}}^\gamma $ are invariant in respect to action of 
\textbf{DT}, and $C_{\rm {\bf x}}^\gamma $ are invariant in respect to 
action of ${\rm {\bf DT}}^{ - 1}$, i.e. for all 
${\rm {\bf x}} \in A \quad {\rm 
{\bf DT}}_{\rm {\bf x}} (S_{\rm {\bf x}}^\gamma ) \subset S_{{\rm {\bf 
T}}({\rm {\bf x}})}^\gamma $ 
and ${\rm {\bf DT}}_{\rm {\bf x}}^{ - 1} 
(C_{\rm {\bf x}}^\gamma ) \subset C_{{\rm {\bf T}}^{ - 1}({\rm {\bf 
x}})}^\gamma $.
\end{enumerate}

If the formulated conditions are valid for a whole absorbing domain 
containing the attractor, say, ${\rm {\bf T}}^n(U)$, they are obviously true 
for the attractor $A = \bigcap\limits_{n = 1}^\infty {{\rm {\bf T}}^n(U)} $.

Let us consider in some detail the procedure of computer verification of 
these conditions. Been given a point 
${\rm {\bf x}} = \left\{ {x_0,\,x_1,\,x_2 ,\,x_3 } \right\} \in U$, 
we perform numerical solution of Eqs. (\ref{eq4}) 
on the interval $t \in \left[ {0,\,\,T} \right]$ 
with the initial state
\begin{equation}
\label{eq8}
\begin{array}{c}
x\vert _{t = 0} = r_0 x_0 ,\,\,u\vert _{t = 0} = r_1 x_1 + c_{ux} x, \\
y\vert _{t = 0} = x_2 + c_{yx} x + c_{yu} u,\,\,v\vert _{t = 0} = x_3 
+ c_{vx} x + c_{vu} u + c_{vy} y
\end{array}
\end{equation}
and get the image
\begin{equation}
\label{eq9}
\begin{array}{ll}
{\rm {\bf {x}'}} = {\rm {\bf T}}({\rm {\bf x}}) &= 
\left\{ {{x}'_0 ,\,{x}'_1,\,{x}'_2 ,{x}'_3 } \right\} \\
&= \{{x}' / r_0 ,\,\,({u}' - c_{ux} {x}') / r_1 
,\,\,{y}' - c_{yx} {x}' - c_{yu} {u}',\,\,v - c_{vx} {x}' - c_{vu} {u}' - 
c_{vy} {y}'\}.
\end{array}
\end{equation}
In parallel, we solve numerically the linearized equations for vectors of 
small perturbations over the same period. In the original variables they are 
\begin{equation}
\label{eq10}
\begin{array}{ll}
\delta \dot {x} = \omega _0 \delta u,\, & \delta \dot {u} = 
- \omega_0 \delta x - 2xu\delta x 
+ (A\cos {2\pi t}/T - x^2)\delta u + 
(\varepsilon /{\omega _0 })\delta y\cos \omega _0 t, \\ 
\delta \dot {y} = 2\omega _0 \tilde {v},\, & \delta \dot {v} = 
- 2\omega_0 \delta y - 2yv\delta y + 
( - A\cos {2\pi t}/ T - y^2)\delta v + 
(\varepsilon / {\omega _0 })x\delta x. \\ 
 \end{array}
\end{equation}
Passage to the redefined coordinates and back may be done with the relations 
\begin{equation}
\label{eq11}
\begin{array}{c}
\delta x_0 = {\delta x}/{r_0 },\,
\delta x_1 = {(\delta u - c_{ux} \delta x)}/{r_1 },\, 
\delta x_2 = \delta y - c_{yx} \delta x - c_{yu} \delta u,\\ 
\delta x_3= \delta v - c_{vx} \delta x - c_{vu} \delta u - c_{vy} \delta y. 
\end{array}
\end{equation}

The equations (\ref{eq10}) are solved along the orbit 
started at \textbf{x} for four 
times, each time with such an initial vector 
\textbf{u}$=\{ \delta x_{i} \}$ that unity 
is posed in a row from 0 to 3, and other elements are 
zero. Then, we get four vector-columns and compose a matrix 
${\rm {\bf U}} = {\rm {\bf DT}}_{\rm {\bf x}} $ of them.

Starting at \textbf{x}, an initial perturbation vector \textbf{u} after one 
iteration of the Poincar\'{e} map yields 
${\rm {\bf {u}'}} = {\rm {\bf Uu}}$. A squared Euclidean norm of this vector is $\left\| {{\rm {\bf u}}'} 
\right\|^2 = {\rm {\bf u}}^T{\rm {\bf U}}^T{\rm {\bf Uu}}$, where $T$ means the 
transposition. Using the inverse matrix ${\rm {\bf U}}^{ - 1}$ we can write 
as well ${\rm {\bf u}} = {\rm {\bf U}}^{ - 1}{\rm {\bf {u}'}}$ and $\left\| 
{\rm {\bf u}} \right\|^2 = {\rm {\bf u}}'^T{\rm {\bf U}}^{ - 1,T}{\rm {\bf 
U}}^{ - 1}{\rm {\bf u}}'$. A condition that ${\rm {\bf {u}'}}$ represents an 
image of a vector belonging to the expanding cone $S_{\rm {\bf x}}^\gamma $, 
is given by an inequality $\left\| {\rm {\bf {u}'}} \right\| > \gamma 
\left\| {\rm {\bf u}} \right\|$, or ${\rm {\bf u}}'^T\left( {{\rm {\bf U}}^{ 
- 1,T}{\rm {\bf U}}^{ - 1} - \gamma ^{ - 2}} \right){\rm {\bf u}}' < 0$. 
Starting at ${\rm {\bf {x}'}} = {\rm {\bf T}}({\rm {\bf x}}) = \left\{ 
{{x}'_0 ,\,{x}'_1 ,\,{x}'_2 ,{x}'_3 } \right\}$, an initial vector ${\rm 
{\bf {u}'}}$ transforms into ${\rm {\bf {u}''}} = {\rm {\bf {U}'{u}'}}$, and 
we have $\left\| {\rm {\bf {u}''}} \right\|^2 = {\rm {\bf {u}'}}^T{\rm {\bf 
{U}'}}^T{\rm {\bf {U}'{u}'}}$. The expanding cone $S_{{\rm {\bf T}}({\rm 
{\bf x}})}^\gamma $ at ${\rm {\bf {x}'}} = {\rm {\bf T}}({\rm {\bf x}})$ is 
determined by an inequality $\left\| {\rm {\bf {u}''}} \right\| > \gamma 
\left\| {\rm {\bf {u}'}} \right\|$, or ${\rm {\bf {u}'}}^T\left( {{\rm {\bf 
{U}'}}^T{\rm {\bf {U}'}} - \gamma ^2} \right){\rm {\bf {u}'}} > 0$.

The $4 \times 4$ matrix ${\rm {\bf U}'}^T{\rm {\bf U}'}$ is positive definite 
and symmetric. Let 
${\rm {\bf d}}_0 ,{\rm {\bf d}}_1 ,{\rm {\bf d}}_2 ,{\rm {\bf d}}_3 $ 
be vectors of the orthonormal basis. The matrix 
${\rm {\bf D}} = 
\left( {{\rm {\bf d}}_0 ,{\rm {\bf d}}_1 ,{\rm {\bf d}}_2 ,{\rm {\bf d}}_3} 
\right)$ transforms the matrix ${\rm {\bf U}'}^T{\rm {\bf U}'}$ to diagonal 
form:
\begin{equation}
\label{eq12}
{\rm {\bf D}}^T{\rm {\bf {U}'}}^T{\rm {\bf {U}'D}} = 
= \{ \Lambda _i^2 \delta_{ij} \}
\end{equation}
Let the eigenvalues be enumerated in decreasing order. As we have one 
expanding and three contracting directions, then, 
$\Lambda _0^2 > 1$ and $\Lambda _{1,2,3}^2 < 1$. 
Now, we suppose that $\gamma $ is selected in 
such way that $\Lambda _0^2 > \gamma ^2$ and 
$\Lambda _{1,2,3}^2 < \gamma ^2$.\footnote{ 
This property is checked in a course of computations at each 
analyzed point of the absorbing domain naturally: its violation would entail 
a non-correct operation of taking a square root of a negative number. The 
inequalities for eigenvalues of the matrix 
${\rm {\bf U}}_{\rm {\bf x}}^T {\rm {\bf U}}_{\rm {\bf x}}$ 
ensure fulfillment of the condition that a sum 
of subsets of the linear vector space (that is a set of all possible linear 
combinations of vectors from the expanding and contracting cones) is the 
full 4D vector space: $S_{\rm {\bf x}}^\gamma + C_{\rm {\bf x}}^\gamma = 
\mathbb{V}$.} Then, in the matrix
\begin{equation}
\label{eq13}
{\rm {\bf D}}^T({\rm {\bf {U}'}}^T{\rm {\bf {U}'}} - \gamma ^2){\rm {\bf D}}
= \{ (\Lambda _i^2 - \gamma ^2) \delta_{ij} \}
\end{equation}
we have one positive and three negative elements on the diagonal. By an 
additional scale change along the coordinate axes 
\begin{equation}
\label{eq14}
{\rm {\bf S}} = \{ s_i^{-1} \delta_{ij} \}, \,\, 
s_0=\sqrt {\Lambda _0^2 - \gamma ^2}, \,\, 
s_{1,2,3}=\sqrt {\gamma ^2 - \Lambda _{1,2,3}^2} 
\end{equation}
it is reduced to the canonical form
\begin{equation}
\label{eq15}
{\rm {\bf S}}^T{\rm {\bf D}}^T({\rm {\bf {U}'}}^T{\rm
{\bf {U}'}} - \gamma ^2){\rm {\bf DS}} = {\rm {\bf {H}'}} = 
\{ h'_i \delta_{ij} \},\,\,h'_0=1,\,\,h'_{1,2,3}=-1.
\end{equation}

A vector \textbf{c}={\{}1,$c_{1}$,$c_{2}$,$c_{3}${\}} 
belongs to 
the expanding cone 
$S_{{\rm {\bf T}}({\rm {\bf x}})}^\gamma $, 
if ${\rm {\bf c}}^T{\rm {\bf {H}'c}} > 0$, or
\begin{equation}
\label{eq16}
c_1^2 + c_2^2 + c_3^2 < 1.
\end{equation}
In the 3D space {\{}$c_{1}$,$c_{2}$,$c_{3}${\}} it corresponds to interior of 
the unit ball.

With the same transformations, the matrix 
${\rm {\bf U}}^{ - 1,T}{\rm {\bf U}}^{ - 1} - \gamma ^{ - 2}$ takes a form
\begin{equation}
\label{eq17}
{\rm {\bf S}}^T{\rm {\bf D}}^T({\rm {\bf U}}^{ - 1,\,T}{\rm {\bf U}}^{ - 1} 
- \gamma ^{ - 2}){\rm {\bf DS}} = {\rm {\bf H}} = \{h_{ij} \}.
\end{equation}
(Note that it is symmetric: $h_{ij} = h_{ji}$.) A vector 
\textbf{c}={\{}1,$c_{1}$,$c_{2}$,$c_{3}${\}} represents an image of a vector 
belonging to the expanding cone, if 
${\rm {\bf c}}^T{\rm {\bf Hc}} < 0$, or
\begin{equation}
\label{eq18}
\begin{array}{l}
h_{00} + h_{01} c_1 + h_{02} c_2 + h_{03} c_3 + h_{10} c_1 + h_{11} c_1^2 + 
h_{12} c_1 c_2 + h_{13} c_1 c_3 + \\ 
 h_{20} c_2 + h_{21} c_1 c_2 + h_{22} c_2^2 + h_{23} c_2 c_3 + h_{30} c_3 + 
h_{31} c_1 c_3 + h_{32} c_2 c_3 + h_{33} c_3^2 < 0. \\ 
\end{array}
\end{equation}
In the space {\{}$c_{1}$,$c_{2}$,$c_{3}${\}} it 
corresponds to interior of some 
ellipsoid. The inclusion ${\rm {\bf DT}}_{\rm {\bf x}} (S_{\rm {\bf 
x}}^\gamma ) \subset S_{{\rm {\bf T}}({\rm {\bf x}})}^\gamma $ means that 
the ellipsoid has to be placed inside the unit ball. To formulate a 
sufficient condition for this, we determine a center of the ellipsoid from 
the equations
\begin{equation}
\label{eq19}
\begin{array}{l}
h_{11} \bar {c}_1 + h_{12} \bar {c}_2 + h_{13} \bar {c}_3 = - h_{10},\\ 
h_{21} \bar {c}_1 + h_{22} \bar {c}_2 + h_{23} \bar {c}_3 = - h_{20},\\
h_{31} \bar {c}_1 + h_{32} \bar {c}_2 + h_{33} \bar {c}_3 = - h_{30},\\
\end{array}
\end{equation}
and estimate a distance of this point from the center of the ball:
\begin{equation}
\label{eq20}
\rho = \sqrt {\bar {c}_1^2 + \bar {c}_2^2 + \bar {c}_3^2 } .
\end{equation}
With a transfer of the origin to the center of the ellipsoid, the equation 
for its surface becomes
\begin{equation}
\label{eq21}
h_{11} \tilde {c}_1^2 + h_{12} \tilde {c}_1 \tilde {c}_2 + h_{13} \tilde 
{c}_1 \tilde {c}_3 + h_{21} \tilde {c}_1 \tilde {c}_2 + h_{22} \tilde 
{c}_2^2 + h_{23} \tilde {c}_2 \tilde {c}_3 + h_{31} \tilde {c}_1 \tilde 
{c}_3 + h_{32} \tilde {c}_2 \tilde {c}_3 + h_{33} \tilde {c}_3^2 = R^2,
\end{equation}
where $\tilde {c}_i = c_i - \bar {c}_i $, and 
\begin{equation}
\label{eq22}
\begin{array}{l}
 R^2 = - (h_{00} + h_{01} \bar {c}_1 + h_{02} \bar {c}_2 + h_{03} \bar {c}_3 
+ h_{10} \bar {c}_1 + h_{11} \bar {c}_1^2 + h_{12} \bar {c}_1 \bar {c}_2 + 
h_{13} \bar {c}_1 \bar {c}_3 + \\ 
 h_{20} \bar {c}_2 + h_{21} \bar {c}_1 \bar {c}_2 + h_{22} \bar {c}_2^2 + 
h_{23} \bar {c}_2 \bar {c}_3 + h_{30} \bar {c}_3 + h_{31} \bar {c}_1 \bar 
{c}_3 + h_{32} \bar {c}_2 \bar {c}_3 + h_{33} \bar {c}_3^2 ). \\ 
 \end{array}
\end{equation}

Now, we consider a symmetric $3\times 3$ matrix 
${\rm {\bf h}} = \{h_{ij} \},\,\,i,j = 1,2,3$. 
In the diagonal representation of this matrix, under 
appropriate orthogonal coordinate transformation 
$(\tilde {c}_1 ,\tilde {c}_2 ,\tilde {c}_3 ) \to (\xi _1 ,\xi _2 ,\xi _3 )$, 
the equation of the ellipsoid surface becomes
\begin{equation}
\label{eq23}
l_1 \xi _1^2 + l_2 \xi _2^2 + l_3 \xi _3^2 = R^2,
\end{equation}
where $l_{1}$, $l_{2}$, $l_{3 }$ are eigenvalues of \textbf{h}. The largest 
semiaxis of this ellipsoid is expressed via the minimal eigenvalue: 
\begin{equation}
\label{eq24}
r_{\max } = R / {\sqrt {l_{\min } } }.
\end{equation}
A sufficient condition for the ellipsoid to be positioned inside the ball is 
given by an inequality
\begin{equation}
\label{eq25}
r_{\max } + \rho < 1.
\end{equation}
It completes the procedure of verification of the expanding cones inclusion 
for the point \textbf{x}.

It may be shown that with $\gamma <$1 the application of the above procedure 
in $U$ is equivalent to verification of the condition in the domain 
\textbf{x}$ \in $\textbf{T}$^{2}(U)$ 
for contracting cones with the parameter 
${\gamma }'= 1 / \gamma > 1$: 
${\rm {\bf DT}}_{\rm {\bf x}}^{ - 1} (C_{\rm {\bf x}}^{1/ \gamma } ) 
\subset C_{{\rm {\bf T}}^{ - 1}({\rm {\bf x}})}^{1 / \gamma }$. 
It is so because the cones $S^\gamma $ and $C^{1 / \gamma }$ are 
complimentary sets: 
$\bar {S}^\gamma \cup \bar {C}^{1 / \gamma } = 
\mathbb{V}$. (Here $S^\gamma $ with $\gamma <1$ corresponds to the cone of 
vectors, which either expand, or contract, but no stronger than by the 
factor $\gamma $.) Hence, fulfillment of the inequality 
(\ref{eq25}) checked inside 
$U$ for two parameters $\gamma $ and $1/\gamma $ would imply that both 
conditions for expanding and for contracting cones are valid in the domain 
\textbf{T}$^{2}(U)$, which contains the attractor.\footnote{ At the same 
$\gamma $, the cones $S^{\gamma }$ and $C^{\gamma }$ have a common border only 
at $\gamma =1$, while at $\gamma >1$ they do not intersect, as required by 
the theorem conditions: 
$S_{\rm {\bf x}}^\gamma \cap C_{\rm {\bf x}}^\gamma = \emptyset $.} 
This is sufficient to draw a conclusion on the hyperbolic 
nature of the attractor. 

The computer verification of the required inclusions for the expanding and 
contracting cones was performed at the parameter values (\ref{eq2}) in the 
coordinate system (\ref{eq4}), (\ref{eq5}). 
Computations of the Poincar\'{e} map and of the 
Jacobi matrices were produced by means of joint numerical solution of the 
differential equations (\ref{eq1}) together with linearized 
equations (\ref{eq10}) on the 
time interval $T$. We used the Runge -- Kutta method of the 8-th order based on 
formulas of Dormand and Prince with automatic selection of step (the 
accuracy for one step was assigned to be 10$^{ - 11})$ and an extrapolation 
method (the accuracy for one step assigned 10$^{ - 15})$ [19]. For solution 
of sets of linear algebraic equations, matrix diagonalization, and 
eigenvalue problem solving, we used sub-programs from the library LAPACK 
[20].

In accordance with our computations, at $\gamma ^{2}=1.1$ the sufficient 
condition (\ref{eq25}) of correct inclusion for the expanding cones 
${\rm {\bf DT}}_{\rm {\bf x}} (S_{\rm {\bf x}}^\gamma ) 
\subset S_{{\rm {\bf T}}({\rm {\bf x}})}^\gamma $ 
is valid in the whole absorbing domain $U$. To discuss 
details, let us consider a 3D hypersurface defined by an equation 
\begin{equation}
\label{eq26}
\left[ {\left( {\sqrt {x_0^2 + x_1^2 } - r} \right) / d_r } \right]^2 + 
\left( {x_2 / d} \right)^2 + \left( {x_3 / d} \right)^2 = R^2.
\end{equation}
At $R=1$ it corresponds to a border of the domain $U$; at $R<1$ 
it belongs to its 
interior. We can parametrize this hypersurface by three angle coordinates 
$\phi $, $\psi $, and $\theta $:
\begin{equation}
\label{eq27}
\begin{array}{ll}
x_0 = (Rd_r \cos \theta + r)\sin \psi,\,\, & 
x_1 = (Rd_r \cos \theta + r)\cos \psi ,\\ 
x_2 = Rd\sin \theta \cos \phi, &
x_3 = Rd\sin \theta \sin \phi.
\end{array}
\end{equation}
The variable $\psi $ may be regarded as a phase of the first oscillator at 
the Poincar\'{e} cross-section, and $\phi $ as a phase of the second 
oscillator at the same instant. Numerical computations on a 3D grid with 
step ${2\pi }/M$ at $M=50$ show that the value 
$r_{\max } + \rho = f(R,\phi ,\psi ,\theta )$ at fixed $R$ 
depends essentially on $\psi $ and 
$\theta $, while dependence on $\phi $ is very weak. 
On the plot of the 
function $f$ one global maximum can be seen of value varied in dependence on 
$\phi $ and $R$. At $R=1$ and some $\phi $ the maximum reaches 
$f_{max} \approx $0.929441 (that corresponds to a point $M$ on the 
border of the domain $U$ with 
coordinates $x_{0} = -0.102628$, $x_{1}= -0.544957$, $x_{2}= 0.000581$, 
$x_{3}= 0.040066$), but remains definitely less than 1, 
see Fig. 3.\footnote{ 
For a search of the maximum in a space of three variables at fixed $R$, we used 
the Newton method.} Panel (b) illustrates mutual disposition for the cones 
${\rm {\bf DT}}_{\rm {\bf x}} (S_{\rm {\bf x}}^\gamma )$ and 
$S_{{\rm {\bf T}}({\rm {\bf x}})}^\gamma $ at the point $M$. 
The plot shows a 3D 
cross-section of the 4D vector space $\mathbb{V} _{\rm {\bf T(x)}}$ 
by a hyperplane 
orthogonal to the expanding direction. The coordinate axes are principal 
semiaxes of the ellipsoid representing the cross-section of the cone 
$S_{{\rm {\bf T}}({\rm {\bf x}})}^\gamma $. Due to scale selection along 
the axes, it looks like a ball. The ellipsoid representing the cross-section 
of ${\rm {\bf DT}}_{\rm {\bf x}} (S_{\rm {\bf x}}^\gamma )$ looks like a 
narrow ``needle'', because of high degree of phase volume compression in two 
directions. Its disposition inside the large ball testifies the condition 
${\rm {\bf DT}}_{\rm {\bf x}} (S_{\rm {\bf x}}^\gamma ) 
\subset S_{{\rm {\bf T}}({\rm {\bf x}})}^\gamma $. 
The ball circumscribed around the ellipsoid is 
posed inside the large ball too; that expresses the sufficient condition 
(\ref{eq25}). For smaller $R$ the global maximum of 
$r_{\max } + \rho $ only decreases 
(Fig. 4a). 
Analogous computations with other values of $\gamma $ indicate 
that the required 
inclusions for the cones $S$ take place at least in the interval 
$0.64< \gamma ^{2}<1.35$ (Fig. 4b). As explained, 
the correctness of the condition with 
$\gamma <1$ implies the condition for the contracting cones 
${\rm {\bf DT}}_{\rm {\bf x}}^{ - 1} (C_{\rm {\bf x}}^{1 / \gamma } ) 
\subset C_{{\rm {\bf T}}^{ - 1}({\rm {\bf x}})}^{1 / \gamma } $ 
in the domain \textbf{T}$^{2}(U)$. We conclude that in 
\textbf{T}$^{2}(U)$, both conditions 
for expanding and contracting cones are true, say, at $\gamma 
^{2}$=1.1.$^{ }$\footnote{ 
Restricting the domain of verification of the 
condition for expanding cones by the set \textbf{T}$^{2}(U)$ one can improve 
essentially the estimate for maximum allowable $\gamma $. In accordance with 
our computations, the inclusion conditions for expanding and contracting 
cones inside the domain \textbf{T}$^{2}(U)$ are valid yet at 
$\gamma ^{2} \approx 1.5$. } 
Hence, the analyzed attractor is uniformly hyperbolic. This 
assertion, although not proven in a classic mathematical style, follows with 
definiteness from the theorem, conditions of which have been checked in the 
computations. Assuming the hyperbolicity established, let us illustrate now 
some attributes of the hyperbolic dynamics.

To start, we note that dynamics on the attractor is chaotic. In a course of 
time evolution, both oscillators generate turn-by-turn, passing the 
excitation one to another. Figure 5 shows typical plots for $x$ and
$y$ obtained from numerical solution of Eqs. (\ref{eq1}). 
Panel (a) presents a single 
sample, and panel (b) shows five superimposed samples of the same signal on 
successive time intervals. Fig. (b) gives evidence that 
the process is not periodic. In fact, it is chaos, which manifests itself in 
irregular displacement of the maxima and minima of the waveforms 
relative to the envelope on successive time intervals $T$.

To have a quantitative indicator of chaos we turn to Lyapunov exponents. 
With multiple iterations of the Poincar\'{e} map and Jacobi matrix 
computations, we trace evolution of four perturbation vectors by means of 
their subsequent multiplication by the Jacobi matrices obtained in a course 
of the evolution. At each iteration, the Gram -- Schmidt orthogonalization 
and normalization are performed for the set of vectors. The Lyapunov 
exponents are determined as the mean rates for growth or decrease of the 
accumulating sums for logarithms of norms for the vectors 
(after orthogonalization but before the normalization) [21]. 
From the computations 
(10 samples, each of $5 \cdot 10^{4}$ iterations of the Poincar\'{e} map) 
we obtained the Lyapunov exponents 
\begin{equation}
\label{eq28}
\begin{array}{ll}
L_1 = 0.6832 \pm 0.0007, & L_2 = - 2.6022 \pm 0.0036, \\
L_3 = - 4.6054 \pm 0.0028, & L_4 = - 6.5381 \pm 0.0078. 
\end{array}
\end{equation}
Presence of the positive exponent $L_{1}$ indicates chaos. (It 
is close to $\ln 2=0.6931$ because of applicability of the 
approximate Bernoulli map (\ref{eq3}).) 

Figure 6 shows portraits of the attractor on a plane of variables of the 
first oscillator. The panel (a) depicts projection of the attractor from the 
5D extended phase space on the plane of original variables $(x,\, u)$. The 
attractor is shown in gray scales (the darkness reflects a relative duration 
of residence inside a given pixel). Black dots relate to the Poincar\'{e} 
cross-section, the instants $t_n = nT$. The panel (b) shows the attractor in 
the Poincar\'{e} cross-section on a plane of the redefined coordinates 
($x_{0},\, x_{1})$ (see (\ref{eq4})). Note an evident visual similarity with the Smale 
-- Williams attractor as depicted in textbooks. The transverse Cantor-like 
structure is illustrated separately on the panels (c) and (d) by magnified 
fragments of the previous picture. 
For quantitative characterization of the fractal structure in the 
Poincar\'{e} cross-section, we have estimated the correlation dimension by 
means of the algorithm of Grassberger and Procaccia. Using a 4-component 
time series ${\rm {\bf x}}_n ={\rm {\bf x}}(t_n)$ obtained
from numerical iterations of the Poincar\'{e} map for $n$=1$\div M$, 
$M$=40000, we get $D$=1.2516$\pm $0.0018 (as a result of averaging over 10 
samples). The dimension estimated from Luapunov exponents with the Kaplan -- 
Yorke formula is $D \approx 1.263$.

From the point of view of theoretical analysis of the hyperbolic attractors, 
one of the principal features is that intersections of local stable and 
unstable manifolds if occur must be transversal. 
In computations, to determine the local manifolds with appropriate accuracy 
one can use the following scheme. Let us have three points on the attractor 
obtained one from another by $N$-fold application of the Poincar\'{e} map: 
${\rm {\bf x}}_A \to {\rm {\bf x}}_B = {\rm {\bf T}}^N({\rm {\bf x}}_A ) \to 
{\rm {\bf x}}_C = {\rm {\bf T}}^N({\rm {\bf x}}_B )$, 
where $N$ is a sufficiently large integer. 
To obtain the 1D unstable manifold at B, we consider an ensemble 
of initial conditions close to $A$ and parametrized by $\Delta \psi $, a small 
deflection of the angle variable, of order $L_1^{ - N} $: 
$x_0 = r^A\sin \psi ,$, 
$x_1 = r^A\cos \psi ,\,\,x_2 = x_2^A ,\,\,x_3 = x_3^A $, 
$r^A = \sqrt {(x_0^A )^2 + (x_1^A )^2} $, 
$\psi = \psi ^A + \Delta \psi $, 
$\psi ^A = \arg (x_1^A + ix_0^A )$. 
After $N$ iterations of the map {\bf T}, the 
points take up positions along the unstable manifold $\Gamma _u^B $. 
To obtain the 3D stable manifold at $B$ 
we set initial conditions for 
the Poincar\'{e} map close to $B$: 
$x_0 = (r^B + \Delta r)\sin \psi _0 ,\,\,
x_1 = (r^B + \Delta r)\cos \psi _0 ,\,\,
x_2 = x_2^B + \Delta x_2 ,\,\,x_3 = x_3^B + \Delta x_3 $, 
where 
$r^B = \sqrt {(x_0^B )^2 + (x_1^B )^2} $. 
Fixing three values ($\Delta r$, $\Delta x_{2}$, $\Delta x_{3})$, which 
parametrize the manifold, we take as initial guess 
$\psi _0 = \psi ^B = \arg (x_1^B + ix_0^B )$ 
and perform $N$ iterations of the map. Then, we get a 
discrepancy 
$\psi _N - \psi ^C$, $\psi _0 = \psi ^C = \arg (x_1^C + ix_0^C )$, 
correct the initial angle variable, 
${\psi }'_0 = \psi _0 +  \quad (\psi ^C - \psi _N ) / 2^N$, 
and repeat the procedure, until the error will be less 
than a given small value. 

A graphic representation of the manifolds is not trivial because the phase 
space is four-dimensional. Let us use a plane of variables 
($x_{0}$, $x_{1})$ relating to the first oscillator. 
The 1D unstable manifold we show 
simply as a projection onto this plane. For 
representation of a 
three-dimensional stable manifold we will use a curve of intersection of the 
manifold with a two-dimensional plane 
$\{x_2 = x_2^B ,\,\,x_3 = x_3^B \}$ 
projected onto the plane ($x_{0}$, $x_{1}$). Practically, a sufficient 
accuracy for coordinates of points on the manifolds is reached, say, at 
$N\sim 10$.
The disposition of the local manifolds revealed from the computations 
is illustrated in Fig. 7. The invariant set that consists of the 
unstable manifolds coincides with the attractor itself. It is enclosed in 
the toroidal absorbing domain going turn by turn around ``the hole of the 
doughnut''. On the other hand, the local stable manifolds are posed across 
the ``tube'' that forms a surface of the toroid. In the 
two-dimensional diagram the stable manifolds look like ``speaks of a 
wheel''. Due to such mutual location, the stable and unstable manifolds can 
intersect only transversally, and no tangencies do occur. 

As stated in this article, in the four-dimensional phase space of 
Poincar\'{e} map for the system of two non-autonomous coupled van der Pol 
oscillators there exists a toroidal absorbing domain, containing a uniformly 
hyperbolic attractor. This conclusion is based on computer verification of 
conditions formulated in terms of appropriate inclusion of expanding and 
contracting cones defined in the tangent vector spaces associated with the 
points of the absorbing domain. Hence, our model delivers a long-time 
expected example of a simple physically realistic system with hyperbolic 
attractor. With this example, it will be possible to construct other models 
with hyperbolic chaos, exploiting structural stability intrinsic to the 
hyperbolic attractors. In fact, a physical experiment demonstrating 
attractor of this type has been performed already on a basis of coupled 
electronic oscillators [16]. In applications, the systems with hyperbolic 
chaos may be of special interest because of their robustness (structural 
stability). An interesting, and now a substantial direction is constructing 
chains, lattices, networks on a base of elements with hyperbolic chaos [22]. 
Models of this class may be of interest for understanding deep and 
fundamental questions, like the problem of turbulence.

\textit{This research was supported by RFBR grant No  06-02-16619.}

\bigskip
\bigskip
\bigskip
\begin{figure}[htbp]
\centerline{\includegraphics[width=5in]{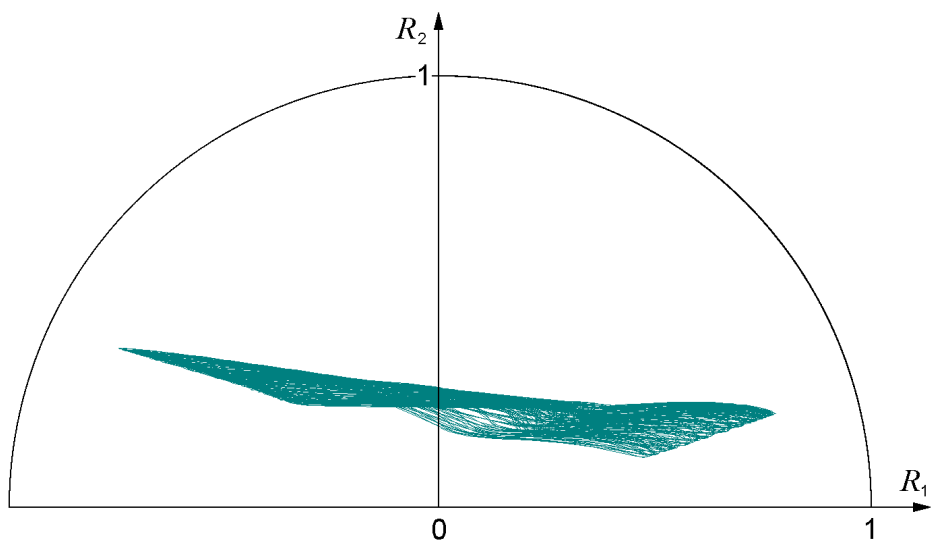}}
\label{fig1}
\caption{
Graphical evidence that the domain $U$ 
defined by (\ref{eq6}) is absorbing. For initial points 
distributed 
over a border of $U$, the resulting data from numerical
solution of the 
differential equations over a period $T$ plotted in 
coordinates (\ref{eq7}) fit inside the unit circle $R_1^2 + R_2^2 = 1$.}
\end{figure}

\begin{figure}[htbp]
\centerline{\includegraphics[width=5in]{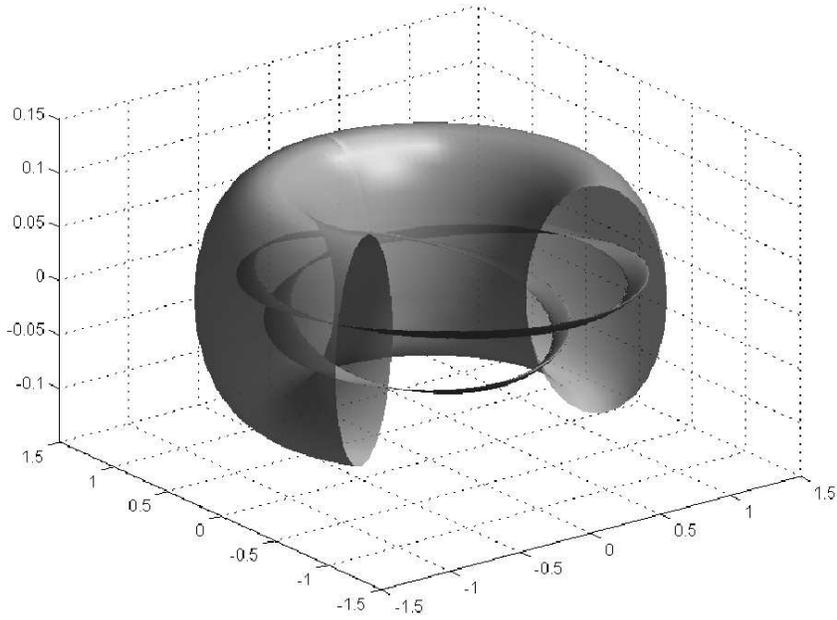}}
\label{fig2}
\caption{
The toroidal absorbing domain $U$ and its image  
{\bf T}$(U)$ shown in a 3D projection. Variables $x_{0}$, 
$x_{1}$ are plotted along the axes in the horizontal plane, and $x_{2}$ along 
the vertical axis. The fourth variable $x_{3}$ corresponds to direction of 
the projecting.}
\end{figure}

\begin{figure}[htbp]
\centerline{
\includegraphics[width=3.6in]{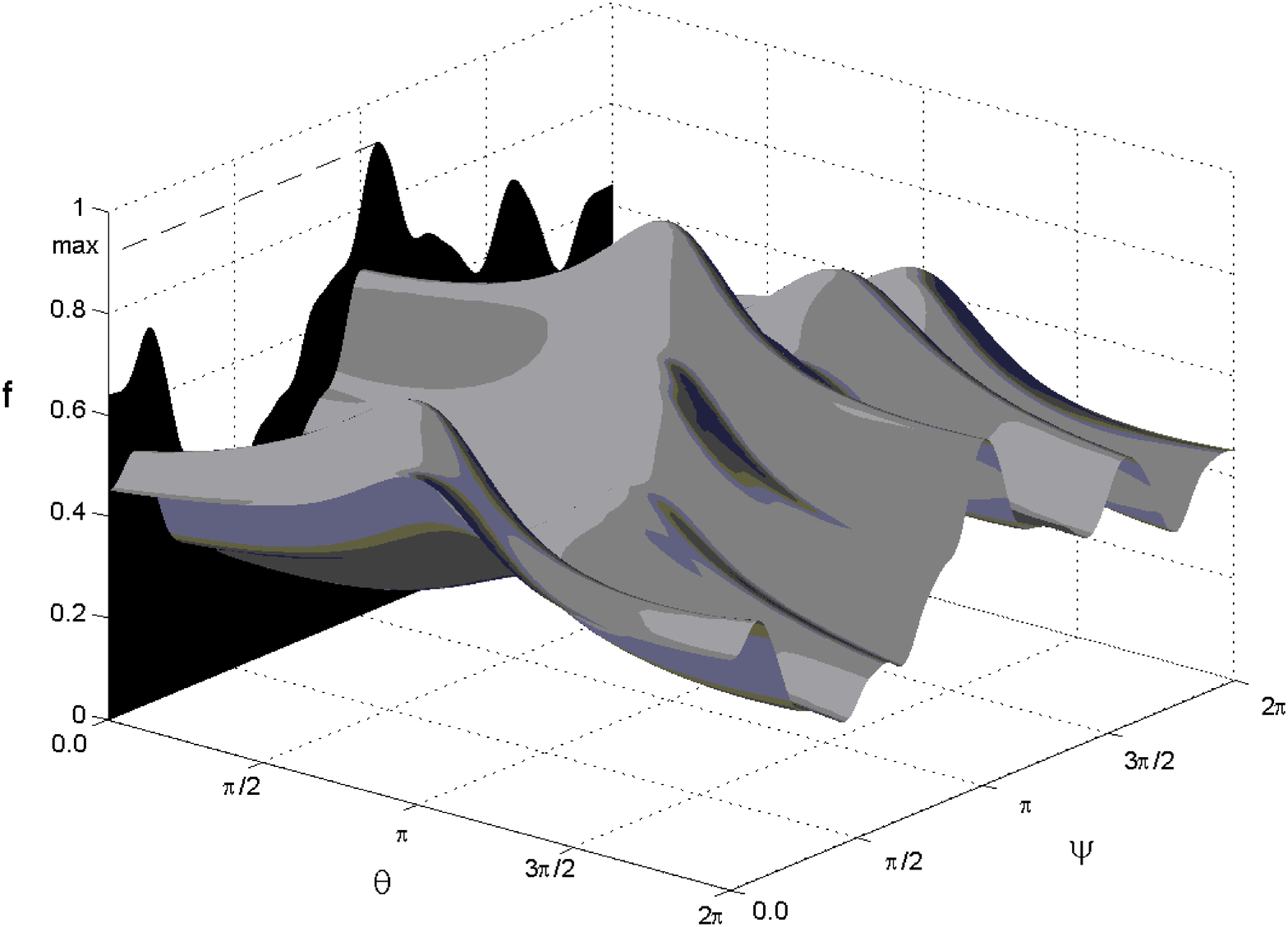}
\includegraphics[width=2.8in]{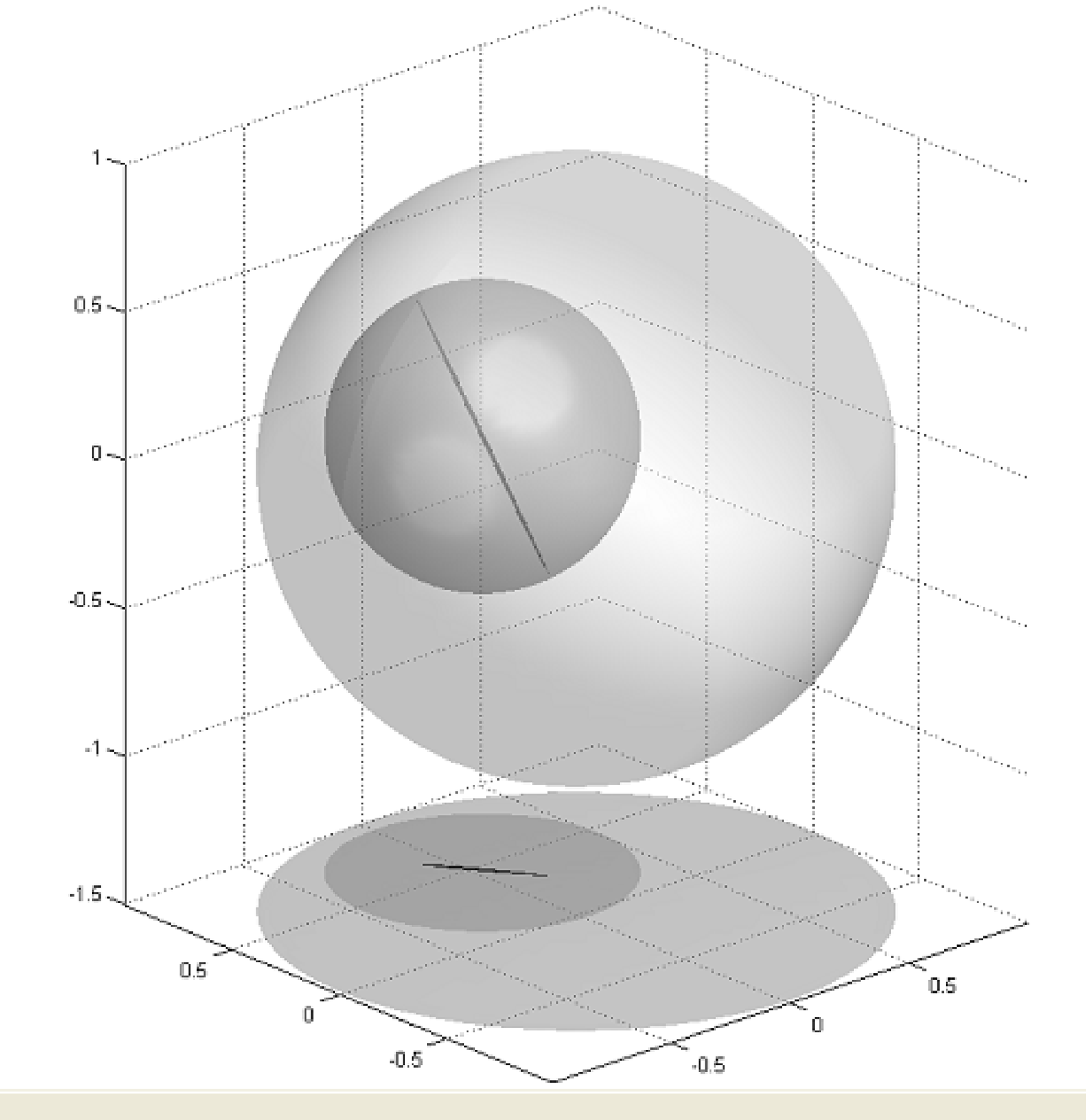}}
\begin{center}
(a)\hspace{3in}(b)
\end{center}
\label{fig3}
\caption{
A plot of the function 
$r_{\max } + \rho = f(R,\phi ,\psi, \theta )$ at $R=1$ and 
$\phi =1.25665$ (a) and a diagram illustrating mutual 
disposition of the 3D cross-sections of the cones 
${\rm {\bf DT}}_{\rm {\bf x}} (S_{\rm {\bf x}}^\gamma )$ and 
$S_{{\rm {\bf T}}({\rm {\bf x}})}^\gamma $ 
at the point of global maximum of $r_{\max } + \rho $. }
\end{figure}

\begin{figure}[htbp]
\centerline{
\includegraphics[height=2.2in]{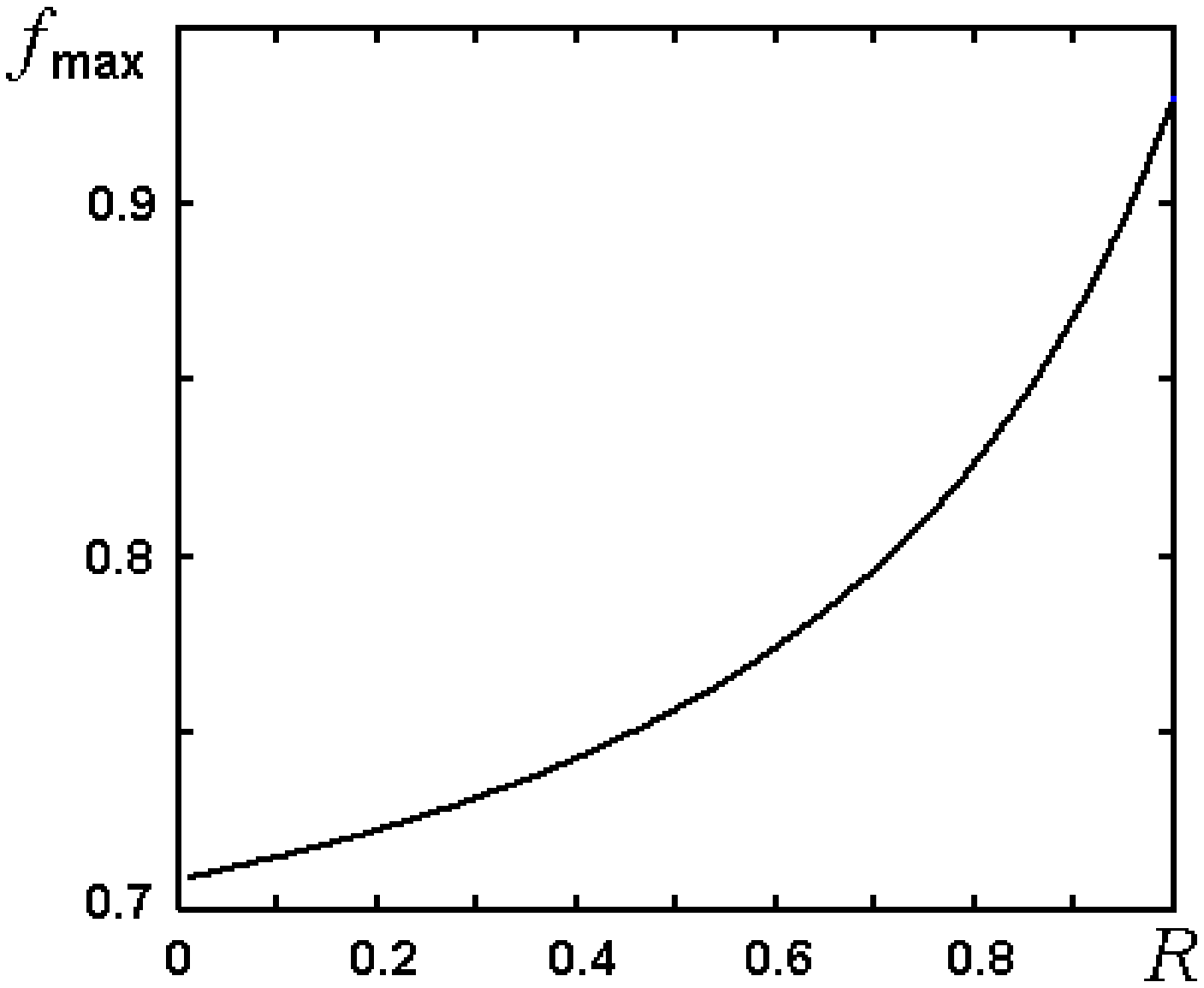}
\includegraphics[height=2.2in]{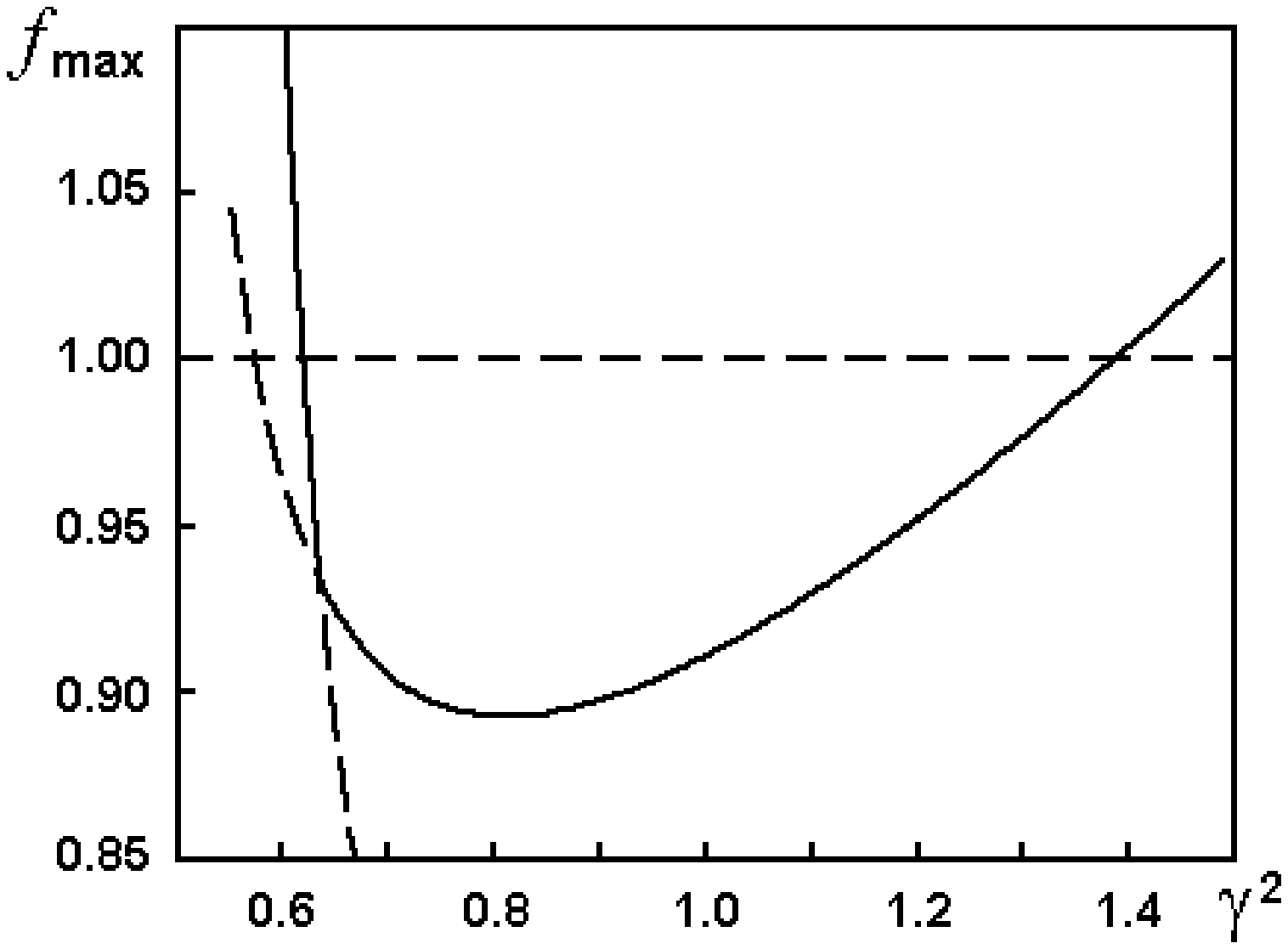}}
\begin{center}
(a)\hspace{3in}(b)
\end{center}
\label{fig4}
\caption{
Value of the global maximum of the function 
$f = r_{\max } + \rho$ on a hypersurface (\ref{eq26}) versus 
parameter $R$ (a) and the global maximum 
value over the whole absorbing domain $U$ 
in dependence on parameter $\gamma $.}
\end{figure}

\begin{figure}[htbp]
\centerline{\includegraphics[width=6.5in]{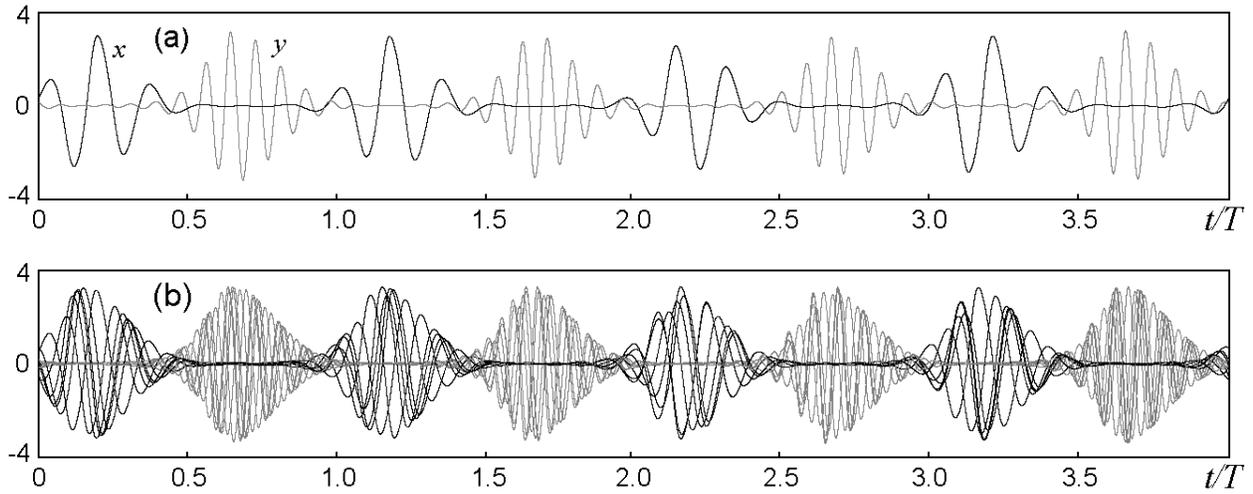}}
\label{fig5}
\caption{
Typical patterns of time dependences for the variables $x$ 
and $y$ obtained from numerical solution of Eqs. (\ref{eq1}) 
at $T = 6$, $A = 5$, 
$\varepsilon = 0.5$. Panel (a) presents a single sample, and panel (b) shows 
five superimposed samples of the same signal on successive time intervals.}
\end{figure}

\begin{figure}[htbp]
\centerline{\includegraphics[width=6.5in]{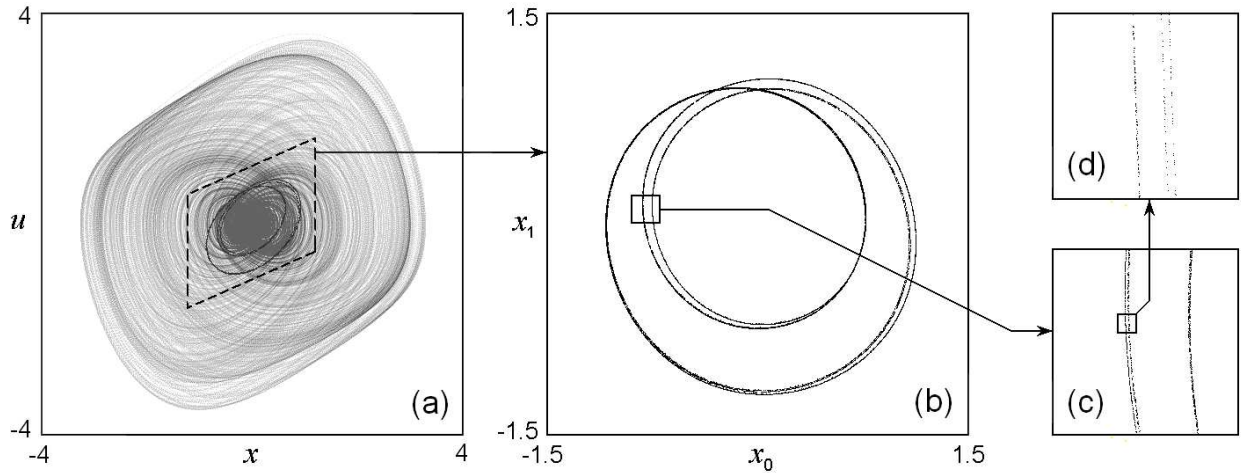}}
\label{fig6}
\caption{
Portraits of the attractor on a plane of variables of the 
first oscillator: (a) -- projection of the attractor from the 5D extended 
phase space on the plane of original variables ($x$, $u)$; 
(b) -- the attractor in 
the Poincar\'{e} cross-section on a plane of the redefined 
coordinates (\ref{eq4}); 
(c), (d) -- details of the Cantor-like transverse structure.}
\end{figure}

\begin{figure}[htbp]
\centerline{\includegraphics[width=4in]{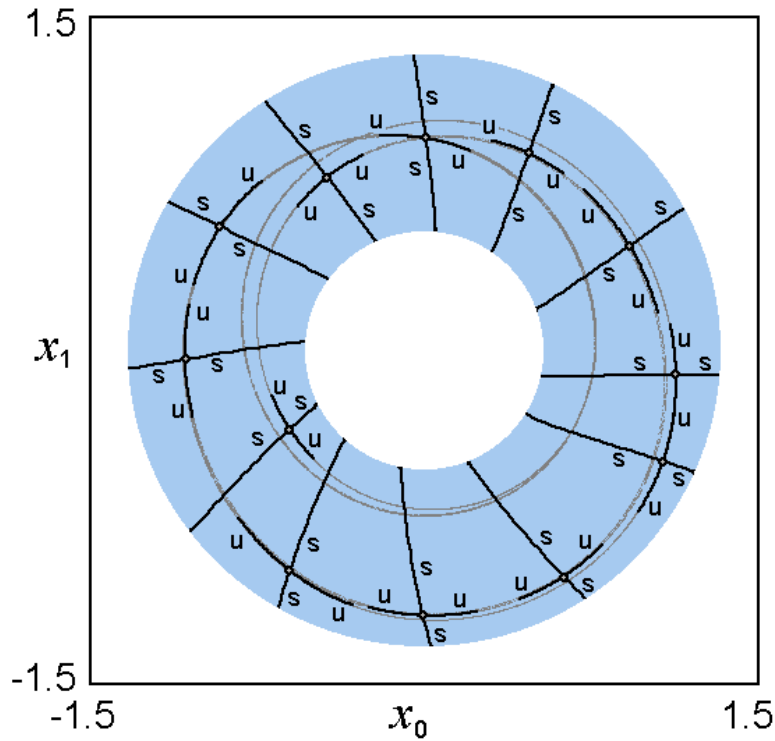}}
\label{fig7}
\caption{
A diagram on the plane of variables $(x_{0}, x_{1})$ 
illustrating mutual location of local unstable (u) and 
stable (s) manifolds for a set of points on the attractor 
in the Poincare cross-section. The gray ring-shaped area 
depicts a projection of the absorbing domain $U$.}
\end{figure}

\end{document}